\newcounter{myctr}
\def\myitem{\refstepcounter{myctr}\bibfont\noindent\ifnum\themyctr>9\else\phantom{0}\fi\hangindent17pt\themyctr.\enskip}

\documentclass{ws-ijqi}
\usepackage{color}

\newcommand{\bra}[1]{\left\langle#1\right\vert}
\newcommand{\ket}[1]{\left\vert#1\right\rangle}
\newcommand{\braket}[2]{\left\langle#1\right\vert\left.#2\right\rangle}
\newcommand{\ketbra}[2]{\left\vert#1\right\rangle\left\langle#2\right\vert}
\newcommand{\h}{\mathcal{H}}

\newcommand{\tr}{\mathrm{tr}}

\def\d{\mathrm{d}}

\begin{document}

\markboth{Cunden, Di Martino, Facchi, Florio}
{Spatial separation and entanglement of identical particles}

%%%%%%%%%%%%%%%%%%%%% Publisher's Area please ignore %%%%%%%%%%%%%%
\catchline{}{}{}{}{}
%%%%%%%%%%%%%%%%%%%%%%%%%%%%%%%%%%%%%%%%%%%%%%%%%%%%%%%%%%%%%%%%%%%

\title{SPATIAL SEPARATION AND ENTANGLEMENT OF IDENTICAL PARTICLES }

\author{FABIO DEELAN CUNDEN}

\address{Dipartimento di Matematica, Universit\`a di Bari, Bari, I-70125, Italy\\
fabio.cunden@uniba.it}

\author{SARA DI MARTINO}

\address{Dipartimento di Matematica, Universit\`a di Bari, Bari, I-70125, Italy\\
sara.dimartino@uniba.it}

\author{PAOLO FACCHI}

\address{Dipartimento di Fisica and MECENAS, Universit\`a di Bari, BAri, I-70126, Italy\\
INFN, Sezione di Bari, Bari, I-70126, Italy\\
paolo.facchi@ba.infn.it}

\author{GIUSEPPE FLORIO}

\address{Museo Storico della Fisica e Centro Studi e Ricerche
``Enrico Fermi'', Piazza del Viminale 1, Roma,  I-00184, Italy\\
Dipartimento di Fisica and MECENAS, Universit\`a di Bari, BAri, I-70126, Italy\\
INFN, Sezione di Bari, Bari, I-70126, Italy\\
giuseppe.florio@ba.infn.it}

\maketitle

\begin{history}
\received{Day Month Year}
\revised{Day Month Year}
%\accepted{Day Month Year}
%\comby{(xxxxxxxxxx)}
\end{history}

\begin{abstract}
We reconsider the effect of indistinguishability on the reduced density operator of the internal degrees of freedom (tracing out the spatial degrees of freedom) for a quantum system composed of identical particles located in different spatial regions. We explicitly show that if the spin measurements are performed in disjoint spatial regions then there are no constraints on the structure of the reduced state of the system. This implies that the statistics of identical particles has no role from the point of view of separability and entanglement when the measurements are spatially separated. We extend the treatment to the case of $n$ particles and show the connection with some recent criteria for separability based on subalgebras of observables.
\end{abstract}


\section{Introduction}\label{sec:intro}

Since the early days of quantum mechanics it has been realized that indinguishability of quantum particles is a fundamental feature of the theory and has important consequences for the interpretation of physical phenomena.\cite{Messiah} From the point of view of the present research activity, it is worth noticing that novel applications in the context of quantum technologies such as sub-shot-noise quantum metrology\cite{Wineland,Benatti2010} rely on the consequences of the symmetrization postulate of quantum mechanics and on the  role of a fundamental resource as quantum entanglement.\cite{Amico,Horodecki} The latest is a direct consequence of linearity (superposition principle) in tensor product Hilbert spaces.

Due to its importance both for the comprehension of the foundations of quantum mechanics and for experimental applications, the study of the relation between correlations and the properties of identical particles has attracted a lot of attention in recent years. 

A careful definition of entanglement criteria for the state vector of a system of identical particles is necessary and can be based on the properties of the Schmidt eigenvalues and on the evaluation of the von Neumann entropy of the single-party reduced density operators.\cite{Ghirardi1,Ghirardi2} Recent results stress the fact that one should also consider the measurement prescription used in the experiment.\cite{Tichy}
The concepts of separability has been extended to the case of identical particles also in terms of commuting algebras of observables\cite{Benatti2011} ensuring the entanglement detection through a partial transposition criterion.\cite{Benatti2012} Moreover, the use of representation theory of the symmetry group can lead to distinguish the entanglement of pure states using a proper generalization of the notion of Schmidt rank.\cite{Marmo} Finally, an approach through the GNS construction\cite{Bratteli} has been recently proposed in Ref.~\refcite{Bala} based on the use of the general idea of the restriction of states to subalgebras.

As a matter of fact, this intense research activity shows that the problem still deserves attention and further analysis in order to be completely clarified. In this paper we will investigate the properties of states of indistinguishable particles. In particular, we will elaborate on an idea introduced by Peres\cite{Peres} and related to the notion of cluster separability. By an explicit calculation (after tracing out  the spatial degrees of freedom), we will show that the condition of spatial separation (described in terms of disjoint domains of the projection operators) makes the role of indistinguishability completely ineffective on the structure of the reduced density operator. This condition has been already recognized as the natural request in order to recover the distinguishability of identical particles in experiments with fermions and bosons.\cite{Herbut1,Herbut2} As a consequence, the standard paradigm of Alice and Bob, largely used in the context of quantum communication and quantum information processing, is still valid in the presence of identical particles. We will also extend this analysis to the case of $n$ identical particles.

The paper is organized as follows. In Section \ref{sec:identical} we will briefly review some fundamental notions and the formalism for describing  indistinguishable particles. In Section \ref{sec:trace} we will show how the partial trace over the spatial degrees of freedom and the hypothesis of spatial separation among the spin measurements allows to treat the reduced state independently of the particle statistics. We will consider both the case of two and $n$ indistinguishable particles. Section \ref{sec:finite} is devoted to some considerations on symmetry. In Section \ref{sec:subalgebra} we will frame our results in terms of  algebras of observables. Finally, in Section \ref{sec:conclusion} we will draw some conclusions.

\section{Indistinguishable Particles}\label{sec:identical}

In order to set the notation, we briefly review the basic concepts and the  formalism for describing identical quantum particles.

Let us consider a quantum system composed of $n$ particles with spins $s_1,s_2,...,s_n$. Pure states of the whole system are associated to unit vectors in the Hilbert space 
\begin{equation}
\mathcal{H}_{\mathrm{total}}=\h_1\otimes\dots\otimes\h_n\ ,
\end{equation}
a tensor product of  single particle Hilbert spaces $\h_k$, with $k=1,\dots,n$. 
Any particle has two relevant sets of degrees of freedom related to its
spin and its position in space. Then, the Hilbert 
space $\h_k$ describing the state of the $k$-th particle is itself the product of a spin
space $\mathfrak{h}_k$ and a spatial space $\mathfrak{l}_k$:
\begin{equation}
\h_k= \mathfrak{l}_k \otimes\mathfrak{h}_k
\end{equation}
If the $k$-th particle has spin $s_k$ and is localized in a region $\Omega_k\subset\mathbb{R}^3$, then $\mathfrak{h}_k=\mathbb{C}^{2 s_k +1}$ is a $2s_k+1$ dimensional complex vector space and $\mathfrak{l}_k=L^2(\Omega_k)$ is the space of square integrable functions on $\Omega_k$. 
A pure state of the system is therefore described by a normalized  wave function $\Psi({\bf x}_1,\sigma_1,...,{\bf x}_n,\sigma_n)$, with ${\bf x}_k\in \Omega_k$ and $\sigma_k=-s_k,-s_k+1,...,s_k$

Now let us focus on systems composed of particles of identical nature. According to the previous setting, for a system of $n$ identical particles (same mass $m$ and spin $s$) we have to consider the $n$-fold tensor product of identical one-particle Hilbert spaces
\begin{equation}\label{eq:identicalspace}
\mathcal{H}_{\mathrm{total}}=\h^{\otimes n}={\left(\mathfrak{l} \otimes\mathfrak{h}\right)}^{\otimes n},
\end{equation}
with $\mathfrak{h}=\mathbb{C}^{2 s +1}$ and, in general, $\mathfrak{l}=L^2(\mathbb{R}^{3})$.
By indistinguishability we mean that any property of the composite system has to be invariant under a relabeling of the particles.
Such indistinguishability of the quantum particles implies a strong limitation on the possible states of the  system: the admissible vectors belong to a proper subspace of $\mathcal{H}_{\mathrm{total}}$. Indeed, let us  consider the action of the symmetric group $S_n$ on $\mathcal{H}_{\mathrm{total}}$
\begin{equation}
W_\pi\Psi({\bf x}_1,\sigma_1,...,{\bf x}_n,\sigma_n)=\Psi({\bf x}_{\pi (1)},\sigma_{\pi (1)},...,{\bf x}_{\pi (n)},\sigma_{\pi (n)}), \;\;\pi\in S_n.
\label{eq:action}
\end{equation}
Such $W_\pi$'s provide a unitary representation of $S_n$ in $\h_n$ in the sense that:
\begin{equation}
W_\pi W_{\tau}=W_{\pi\tau}\quad \mathrm{and}\quad W_{\pi}^{\dagger}=W_{\pi}^{-1}=W_{\pi^{-1}}\,,\quad\forall \pi,\tau\in S_n.
\label{eq:unitary_repr}
\end{equation}
A relabeling of the $n$ particles according to $\pi\in S_n$, trasforms the state of the system $\Psi\mapsto W_\pi\Psi$ according to Eq.\ (\ref{eq:action}).
It turns out that, for $n$ identical particles, all unitaries $W_\pi$'s must leave the state $\Psi$ unchanged  apart from a constant, $W_\pi\Psi=\lambda \Psi$, and in $\mathbb{R}^3$ there are only two possible values $\lambda=\pm1$ that are consistent with the linearity of quantum mechanics\cite{Messiah}. The vector states describing $n$ identical bosons belong  to the symmetric subspace
\begin{equation}
\h^{\odot n}=\{ \Psi\in \h^{\otimes n}: W_\pi\Psi=\Psi, \forall \pi\in S_n\},
\end{equation}
while those describing $n$ identical fermions belong to the antisymmetric subspace
\begin{equation}
\h^{\wedge n}=\{\Psi\in \h^{\otimes n}:W_\pi\Psi=\mathrm{sgn}({\pi})\Psi, \forall \pi\in S_n\}.
\end{equation}
Notice that the orthogonal projections\cite{Bratteli}
\begin{equation}
\Pi_+
=\frac{1}{n!}\sum_{\pi\in S_n}{W_\pi},\qquad\mathrm{and}\quad 
\Pi_- =\frac{1}{n!}\sum_{\pi\in S_n}{\mathrm{sgn}(\pi) W_\pi}
\label{eq:orth_proj_2}
\end{equation} 
map onto the above subspaces:
\begin{equation}
\h^{\odot n}=\Pi_+
\mathcal{H}_{\mathrm{total}},\qquad \mathrm{and}\quad 
\textstyle\h^{\wedge n}=\Pi_-
\mathcal{H}_{\mathrm{total}}\ .
\end{equation}

Finally, let us spend a few words on the appropriate algebra of observables of system of  $n$ identical particles. The outcomes of a measurement of an $n$-particle state $\rho$ have to be invariant with respect to any permutation on the state $W_\pi\rho W_\pi^{\dagger}$. This condition imposes that not all operators in $\mathcal{B}(\h^{\otimes n})$ (the algebra of all bounded operators acting on $\h^{\otimes n}$) are  appropriate observables of the system. To be more precise, the above requirement imposes that for identical particles, an observable $X$ must lie on the subalgebra of the \emph{exchangeable} operators:
\begin{equation}
X=W_\pi XW_\pi^{\dagger}\ .
\label{eq:exchangeable}
\end{equation}
A one-particle operator $A\in\mathcal{B}(\h)$ is lifted in a natural way to an observable of $n$ identical particles by taking into account  condition (\ref{eq:exchangeable}). Such a lift is  provided by the map (second-quantization functor)  
$\d\Gamma:\mathcal{B}(\h)\mapsto\mathcal{B}(\h^{\otimes n})$,
whose action is\cite{Bratteli}
\begin{equation}
\label{eq:gamma}
\d\Gamma(A)= A\otimes  \mathbb{I}\otimes\dots\otimes  \mathbb{I}+\mathbb{I}\otimes A\otimes\dots\otimes  \mathbb{I}+\dots+ \mathbb{I}\otimes\dots\otimes  \mathbb{I}\otimes A,
\end{equation}
where $\mathbb{I}$ is the identity operator on $\h$. 
It is easy to see that the symmetric and antisymmetric subspaces are left invariant by the operator $\d\Gamma(A)$.

More generally, the second quantization functor acts in a natural way on any $k$-particle operator, with $k\leq n$.
In this paper we will make use only of the second quantization of  $n$-particle operators whose explicit expression is
\begin{equation}
\d\Gamma(A_1\otimes A_2\otimes \dots \otimes A_n) = 
\sum_{\pi\in S_n} A_{\pi(1)}\otimes A_{\pi(2)}\otimes \dots \otimes A_{\pi(n)}
\label{eq:gamman}
\end{equation}
As a particular case, consider
\begin{equation}
\Pi_{\mathrm{spatial}}= \d\Gamma\Big((P_1\otimes\mathbb{I})\dots\otimes (P_n\otimes\mathbb{I}) \Big)
\label{eq:Big_Proj}
\end{equation}
where the $P_i$'s act on the spin space $\mathfrak{l}$, while the identity operators act on the  spatial space $\mathfrak{h}$. It is easy to verify by inspection that $\Pi_{\mathrm{spatial}}$ is a projection operator provided that the $P_i$'s are orthogonal projections $P_iP_j=\delta_{ij}P_i$.

\section{Partial Trace of the Spatial Degrees of Freedom}\label{sec:trace}

In this section we will show that it is possible to obtain a separable state from a global state describing a system of identical particles under very natural hypotheses. In particular, we will see that it is sufficient to consider projections for the spatial degrees of freedom with disjoint domains (i.e. spatially separated particles) and perform a partial trace in order to obtain a separable quantum state  in the spin degrees of freedom. In order to make a first acquaintance with this problem, we first analyze the case of a system composed by two particles.\cite{Herbut1,Herbut2} 

Let us recall Peres' approach to the analysis of entanglement of identical particles\cite{Peres}, based on the notion of \emph{cluster separability}. Consider two states $\ket{u}$ and $\ket{v}$ in $\h$. We will say that  state $\ket{v}$ is \emph{remote} with respect to $\ket{u}$ if $\|Av\|$ is vanishingly small for any  operator $A$ with support in a spatial neighborhood of $\ket{u}$. It follows that any matrix element of $A$ involving $\ket{v}$ is vanishingly small.
Let us consider a state of two identical particles  
\begin{equation}
\ket{\Psi}=(\ket{u}\otimes\ket{v}\pm\ket{v}\otimes\ket{u} )/\sqrt{2},
\label{eq:entangled}
\end{equation}
where the orthogonal states $\ket{u}$ and $\ket{v}$ describe two particles that are far apart. The state of the pair is entangled but this entanglement has no effect if we focus on localized observables. 
Suppose that the one-particle operator $A\in\mathcal{B}(\h)$ is nonvanishing in a neighborhood  of $\ket{u}$. As a consequence $\|Av\|$ is vanishingly small and then
its lifting $\d\Gamma(A)=A\otimes\mathbb{I}+\mathbb{I}\otimes A$ yields
\begin{equation}
\bra{\Psi}\d\Gamma(A)\ket{\Psi}=\bra{u}A\ket{u}.
\label{eq:expectation}
\end{equation}
because the terms involving $\ket{v}$ will vanish. It is now clear that the requirement of spatial separation is the key ingredient underlying the notion of cluster separability. 

Let us elaborate on this and focus on a system of two identical particles that have both spatial and spin degrees of freedom. 
As  stated in Section \ref{sec:identical}, the one-particle Hilbert space is
\begin{equation}
\h=\mathfrak{l}\otimes \mathfrak{h} = L^2(\mathbb{R}^3)\otimes\mathbb{C}^{2s+1}\ ,
\end{equation}
while the Hilbert space  for a system of two bosons (fermions) is $\h^{\odot 2}$ ($\h^{\wedge 2}$).

Since we are not dealing with product spaces, we will start from the very  definition of the reduced state $\rho_{\mathrm{spin}}$ (of  the spin degrees of freedom):
\begin{equation}
\label{eq:spatial_trace}
\tr (\rho_{\mathrm{spin}}\, A\otimes B)= \tr \bigg\{ \rho\, \d\Gamma\bigg((P\otimes A)\otimes (Q\otimes B)\bigg)\bigg\},
\qquad \forall A,B\in\mathcal{B}(\mathfrak{h}) ,
\end{equation}
where $\rho$ is a generic state in $\h^{\odot 2}$ (or $\h^{\wedge 2}$), $A$, $B$ are two observables acting on spins and $P$, $Q$ are projections  onto the spatial regions  $\Omega_1$ and $\Omega_2$ of $\mathbb{R}^3$, where the two spin measurements  are  respectively performed. Notice that the reduced state will in general depend on the choice of the projections $P$ and $Q$.

The right hand side of Eq.~(\ref{eq:spatial_trace})  is the sum of two terms 
\begin{eqnarray}
\nonumber
& &\tr \left\{ \rho\, \d\Gamma\Big((P\otimes A)\otimes (Q\otimes B)\Big)\right\}\\
& &\qquad =\tr \left\{ \rho\, (P\otimes A)\otimes (Q\otimes B)\right\}+\tr \left\{ \rho\, (Q\otimes B)\otimes (P\otimes A)\right\}.
\label{eq:reducedspin}
\end{eqnarray}
The first term can be recast in the form
\begin{equation}
\label{eq:trace}
\tr \left\{ \rho\, (P\otimes A)\otimes (Q\otimes B)\right\}=\tr \left\{(P\otimes\mathbb{I})\otimes (Q\otimes \mathbb{I}) \rho\, (P\otimes A)\otimes (Q\otimes B)\right\},
\end{equation}
and analogously for the second.

Let us consider  a pure state $\rho=\ketbra{\psi}{\psi}$ with
\begin{equation}
\psi=\frac{1}{\sqrt{2}}\bigg[(f\otimes \xi)\otimes (g\otimes \eta)\pm (g\otimes \eta)\otimes (f\otimes \xi)\bigg],
\label{eq:state_factor}
\end{equation}
where $\xi$, $\eta$ represent one-particle spin states  and $f$, $g$ are one-particle spatial wavefunctions such that $P\, f=f$ and $Q\, g=g$. 

Here we suppose that the two spin measurements are located in disjoint regions of the space, $\Omega_1\cap\Omega_2=\emptyset$ so that $PQ=0$. This is in agreement with the standard setting of quantum communication, where  Alice and Bob are spatially separated, whence the wavefunction $f$ vanishes outside $\Omega_1$ and, similarly,  $g$ is zero outside $\Omega_2$. 

Plugging this state into Eq.~(\ref{eq:reducedspin}) we obtain
\begin{eqnarray}
\nonumber
& &\tr \left\{ \rho\, \d\Gamma\Big((P\otimes A)\otimes (Q\otimes B)\Big)\right\}\\\nonumber
& &\qquad = \frac{1}{2}\tr \left\{ (\ketbra{\xi}{\xi}\otimes\ketbra{\eta}{\eta})( A\otimes  B)\right\}+\frac{1}{2}\tr \left\{ (\ketbra{\eta}{\eta}\otimes\ketbra{\xi}{\xi})( B\otimes  A)\right\}\\
& &\qquad =\tr \left\{ (\ketbra{\xi}{\xi}\otimes\ketbra{\eta}{\eta})( A\otimes  B)\right\}.
\end{eqnarray}
Therefore, from definition~(\ref{eq:spatial_trace}) one finally gets
\begin{equation}
\rho_{\mathrm{spin}}=\ketbra{\xi}{\xi}\otimes\ketbra{\eta}{\eta},
\end{equation}
a pure separable state of the two spins.

In the same way we can start from a generic  symmetric or antisymmetric state (we do not care about normalization)
\begin{equation}
\psi=\frac{1}{\sqrt{2N}}\sum_{i=1}^N \bigg[(f_i\otimes \xi_i)\otimes (g_i\otimes \eta_i)\pm (g_i\otimes \eta_i)\otimes (f_i\otimes \xi_i)\bigg],
\end{equation} 
and obtain the state
\begin{equation}
\rho_{\mathrm{spin}}=\frac{1}{N}\sum_{i,j}\Big(\braket{f_i}{f_j}\braket{g_i}{g_j}\Big) \ketbra{\xi_i}{\xi_j}\otimes\ketbra{\eta_i}{\eta_j},
\end{equation}
which is, in general, mixed and entangled.

We conclude that the structure of $\rho_{\mathrm{spin}}$ is not constrained to satisfy any particular symmetry and  it is a generic state of two distinguishable spins in $\mathbb{C}^{2s+1}\otimes \mathbb{C}^{2s+1}$.
Therefore, the reduced state has no memory of the antisymmetric or symmetric structure of the initial state (where  also  the spatial degrees of freedom are considered) as long as the spin measurements are spatially separated.\cite{Herbut1} Under this assumption and considering the spin degrees of freedom, the definition of entanglement for identical indistinguishable particles is not different from the case of distinguishable particles. Of course, if the particles are localized in the same region, $PQ\neq0$, these results are no longer true.

These results can be put in  Peres' framework described at the beginning of this Section.
Indeed, let us consider the state~(\ref{eq:state_factor}) of two localized identical particles (bosons or fermions)
with $\mathrm{supp}f\, \subset\Omega_1$, $\mathrm{supp}\, g\subset\Omega_2$, and $\Omega_1\cap\Omega_2=\emptyset$. The projection operators $P$ and $Q$ onto $\Omega_1$ and $\Omega_2$, respectively, are mutually orthogonal $PQ=0$. In terms of the above notations,  state $g\otimes\eta$ is remote (in the sense previously defined) with respect to  state $f\otimes\xi$ and then, for any operator $A\in \mathcal{B}(\mathbb{C}^{2s+1})$ we have
\begin{equation}
\bra{\Psi}{\d\Gamma(P\otimes A\otimes Q\otimes \mathbb{I})}\ket{\Psi}=\bra{f\otimes\xi}{\mathbb{I}\otimes A}\ket{f\otimes\xi}.
\label{eq:expectation2}
\end{equation}
Therefore, any question about the statistics of the pair is immaterial at the level of the internal (spin) degrees of freedom provided the two systems are spatially separated. 

The same result can be generalized to  a system of $n$ identical particles. Indeed we can take a state $\Psi$ belonging to $\h^{\odot n}$ for bosons or to $\h^{\wedge n}$ for fermions and trace over the spatial degrees of freedom in the way we did for two particles. This means that we have to consider $n$ projections $P_1,\dots, P_n$  projecting on disjoint spatial regions $\Omega_1,\dots,\Omega_n$, corresponding to $n$ spatially separated spin experiments. Thus,
\begin{equation}
\label{eq:spatial_trace2}
\tr \bigg(\rho_{\mathrm{spin}}\, \bigotimes_{i=1}^{n} A_i\bigg)
=\tr \bigg\{ \rho\, \d\Gamma \Big( \bigotimes_{i=1}^{n} P_{i}\otimes A_{i}\Big)\bigg\} .
\end{equation}

Again we suppose that $\rho$ is a pure state $\rho=\ketbra{\psi}{\psi}$ with
\begin{equation}
\psi=\frac{1}{\sqrt{n!}}\sum_{\pi\in S_n} \epsilon_\pi \bigotimes_{i=1}^n\left(f_{\pi(i)}\otimes\xi_{\pi(i)}\right),
\end{equation}
where $\epsilon_\pi=1$ for bosons and $\epsilon_\pi=\mathrm{sgn}(\pi)$  for fermions. This leads to
\begin{eqnarray}
& & \tr \bigg\{ \rho\, \d\Gamma \Big( \bigotimes_{i=1}^{n} P_{i}\otimes A_{i}\Big)\bigg\} =\sum_{\pi\in S_n}\tr \bigg\{\bigotimes_{i=1}^n(P_{\pi(i)}\otimes \mathbb{I})\ketbra{\psi}{\psi} (P_{\pi(i)}\otimes A_{\pi(i)})\bigg\}\nonumber\\
& &\qquad = \frac{1}{n!} \sum_{\pi\in S_n} \tr \bigg\{\bigotimes_{i=1}^n\Big(\ketbra{f_{\pi(i)}\otimes\xi_{\pi(i)}}{f_{\pi(i)}\otimes\xi_{\pi(i)}}A_{\pi(i)}\Big)\bigg\},
\end{eqnarray}
so that
\begin{equation}
\rho_{\mathrm{spin}}=\bigotimes_{i=1}^n\ketbra{\xi_i}{\xi_i}.
\end{equation}
Whence one arrives to the same conclusion as for two indistinguishable particles.

\section{Some considerations}\label{sec:finite}
The aim of this section is to better clarify the interplay between indistinguishability of the global wavefunction and  entanglement of  local states. 
 Recall that the one-particle Hilbert space is a product  $\h=\mathfrak{l} \otimes\mathfrak{h}$, where $\mathfrak{h}$ is the spin space and $\mathfrak{l}$ the position space. In the previous section we have seen that, whenever some localization precludes  the identical particles to share the same spatial state, there is no obstruction on the structure of the reduced state of the internal (spin) degrees of freedom. The spatial separation makes the reduced states insensitive to the quantum statistics of the global state. 
 
 To be more clear, let us see what happens if the particles share the same spatial state. Let us focus on the fermionic setting $\h^{\wedge n}$. It will be useful to introduce an orthonormal basis $\left\{\ket{\mu}\right\}_{\mu\geq1}$ of $\mathfrak{l}$. A state $\ket{\psi}\in\h^{\wedge n}$ whose $n$ factors share the same spatial state belongs to 
\begin{equation}
\h'=\bigoplus_{\mu\geq 1}
(\ket{\mu} \otimes\mathfrak{h})^{\wedge n}=\mathfrak{l}\otimes\left(\mathfrak{h}^{\wedge n}\right)
\label{eq:subspace1}
\end{equation}
which can be considered as a proper subspace of $\h^{\wedge n}$. The partial trace of a state $\rho\in\mathcal{D}(\h')$ acting on such a space provides a reduced state with fermionic character:
\begin{equation}
\mathrm{tr}_{\mathfrak{l}}\mathcal{D}(\h')=\mathcal{D}(\mathfrak{h}^{\wedge n})\ .
\label{eq:subspace1total}
\end{equation}
Note that Eq. (\ref{eq:subspace1total}) holds also if we consider the subspace, describing a symmetric spatial wavefunction,
\begin{equation}
\h''=\left(\mathfrak{l}^{\odot n}\right)\otimes\left(\mathfrak{h}^{\wedge n}\right), 
\qquad  \mathrm{tr}_{\mathfrak{l}^{\odot n}}\mathcal{D}(\h'')=\mathcal{D}(\mathfrak{h}^{\wedge n})\ .
\label{eq:subspace2}
\end{equation} 
On the other hand, if the spatial part of the state  is antisymmetric,  the proper subspace
\begin{equation}
\h'''=\left(\mathfrak{l}^{\wedge n}\right)\otimes\left(\mathfrak{h}^{\odot n}\right)
\label{eq:subspace3}
\end{equation}
is mapped by the partial trace onto the n-fold symmetric product of $\mathfrak{h}$:
\begin{equation}
\mathrm{tr}_{\mathfrak{l}^{\wedge n}} \mathcal{D}(\h''')=\mathcal{D}(\mathfrak{h}^{\odot n})\ .
\label{eq:subspace3total}
\end{equation}
The above three examples should be enough to understand that the statistics of the global state of $n$ identical particles can be hidden, preserved or even changed if we focus on the local states of some very particular subspaces.

\section{Entanglement  and Subalgebras of Observables} \label{sec:subalgebra}
In this section we will show that our results  fit well in other frameworks.
In particular we will show that the main claim of the paper is consistent with
recent findings related to the definition of separability in the context of identical particles. 

Benatti \emph{et al.} (see Ref.~\refcite{Benatti2010,Benatti2011,Benatti2012}) have given the following definition of separability in terms of commuting subalgebras of observables.  Consider $\mathcal{B}(\h)$, the algebra of all bounded operators acting on $\h$. A pair $(\mathcal{A}_1,\mathcal{A}_2)$ of commuting unital subalgebras of $\mathcal{B}(\h)$ is defined to be an algebraic bipartition of $\mathcal{B}(\h)$.

A state $\rho$ on $\mathcal{B}(\h)$ is defined to be separable with respect to the bipartition  $(\mathcal{A}_1,\mathcal{A}_2)$ if for any operator of the form $A_1 A_2$, with $A_1$ belonging to $\mathcal{A}_1$ and $A_2$ belonging to $\mathcal{A}_2$, we have
\begin{equation}
\rho(A_1A_2)= \sum_k \lambda_k \sigma_k(A_1)\omega_k (A_2), \quad \mathrm{with}\;\; \lambda_k\geq 0\;\; \mathrm{and}\;\;\sum_k\lambda_k=1,
\end{equation}
where $\sigma_k$ and $\omega_k$ are states on $\mathcal{B}(\h)$.

We will show that the requirement of spatial separation of  identical particles naturally leads to the bipartition of the algebra of bounded operators, necessary to give a consistent definition of separability.
For the sake of clarity, let us consider two fermions and the algebra of all bounded operators on $\h^{\wedge 2}$, with $\h= \mathfrak{l}\otimes\mathfrak{h}$. The subalgebra of operators concerning the spin degrees of freedom is  
\begin{equation}
\mathcal{A} =\left\{\d\Gamma(\mathbb{I}\otimes A)\,|\, A \in \mathcal{B}(\mathfrak{h})\right\}.
\label{eq:subalg_spin}
\end{equation}
As in Section \ref{sec:trace}, we take two projections  $P,Q$ acting on the spatial space  $\mathfrak{l}$.
Let us consider then the subalgebras of $\mathcal{B}(\h^{\wedge 2})$ defined by: 
\begin{eqnarray}
\label{eq:localized_subalg}
\mathcal{A}_1&=&\left\{\d \Gamma\Big((P\otimes A_1)\otimes (Q\otimes \mathbb{I})\Big)\, |\, A_1\in \mathcal{B}(\mathfrak{h})\right\},\\
\mathcal{A}_2&=&\left\{\d \Gamma\Big((P\otimes \mathbb{I}) \otimes (Q\otimes A_2)\Big)\,|\,A_2\in  \mathcal{B}(\mathfrak{h})\right\}.
\end{eqnarray}
The subalgebras $\mathcal{A}_1,\mathcal{A}_2$ are physically the local subalgebras of (\ref{eq:subalg_spin}) corresponding to spin measurements by Alice and Bob, respectively.

Our claim is that 
subalgebras $\mathcal{A}_1$ and $\mathcal{A}_2$ commutes iff the domains of the projections $P$ and $Q$ do not overlap, namely
\begin{equation}
[\mathcal{A}_1,\mathcal{A}_2]=0 \qquad\mathrm{iff} \qquad PQ=QP=0.
\end{equation}
Indeed, for $A_1,A_2\in\mathcal{B}(\mathfrak{h})$:
\begin{eqnarray}
\nonumber
&&\left[\d\Gamma\Big((P\otimes A_1)\otimes (Q\otimes \mathbb{I})\Big),\d\Gamma\Big((P\otimes \mathbb{I}) \otimes (Q\otimes A_2)\Big)\right]\nonumber\\
& & =[P\otimes A_1\otimes Q\otimes \mathbb{I}+Q\otimes \mathbb{I} \otimes P\otimes A_1 ,P\otimes \mathbb{I}\otimes Q\otimes A_2]\nonumber\\
& & \quad + [ P\otimes A_1\otimes Q\otimes \mathbb{I}+Q\otimes \mathbb{I} \otimes P\otimes A_1,Q\otimes A_2 \otimes P\otimes \mathbb{I}]\nonumber\\
& & = QP \otimes\mathbb{I} \otimes PQ \otimes A_1A_2 - 
 PQ \otimes\mathbb{I} \otimes QP \otimes A_2A_1
\nonumber\\
& & \quad + PQ\otimes A_1A_2\otimes QP \otimes \mathbb{I}
- QP\otimes A_2A_1\otimes PQ \otimes \mathbb{I} .
\end{eqnarray} 
By inspection, this commutator vanishes for any choice of $A_1$ and $A_2$ if and only if $PQ=0$, that is iff the particles are spatially separated.

The projection operators $P$ and $Q$ onto $\Omega_1$ and $\Omega_2$, respectively, are mutually orthogonal $PQ=0$. Then, the subalgebras $\mathcal{A}_1$  of Eq. (\ref{eq:localized_subalg}), corresponds to operators $A_1$ in the spin space for particles localized in $\Omega_1$, and similarly for $\mathcal{A}_2$. 
This result is consistent with the discussion in Section \ref{sec:trace}:  if two systems are spatially separated the statistics has no effect at the level of  the internal degrees of freedom.

\section{Conclusions}\label{sec:conclusion}
We have investigated the fate of spin entanglement in the case where spin is accompanied with spatial degrees of freedom in describing systems of identical particles. We have explicitly shown that, when the spin measurements are performed in disjoint spatial regions, there are no constraints on the structure of the reduced state of the system. Therefore, any question about statistics of identical particles is immaterial at the level of the internal (spin) degrees of freedom provided the particles are spatially separated. We have also shown the connection between our results and some recent criteria for separability based on subalgebras of observables.

\section*{Acknowledgments}

This work is partially supported by PRIN 2010LLKJBX.
The authors acknowledge support from the University of
Bari through the Project IDEA.

\end{document}